
\input phyzzx.tex
\input epsf.tex
\input tables.tex
\overfullrule=0pt
\nopagenumbers
\parskip = 4pt plus 1pt minus 1pt

\def\crr#1{\crcr\noalign{\vskip #1}}
\def\papers{\papersize\headline=\paperheadline\footline=\paperfootline}
\def\papersize{\hsize=17truecm \vsize=25truecm
               \hoffset=-6truemm \voffset=-11truemm}
\papers
\def\refout{\par\penalty-400\vskip\chapterskip
   \spacecheck\referenceminspace
   \ifreferenceopen \Closeout\referencewrite \referenceopenfalse \fi
   \line{\fourteenrm\hfil References\hfil}\vskip\headskip
   \input \jobname.refs}
\def\cap#1{{\leftskip 0.5truein \rightskip 0.5truein \noindent #1 \par}}
\def\propline{\hrule width60pt depth0.5pt height0.5pt}
\def\prop{\vcenter{\propline}}
\def\heavyprop{\vcenter{\propline\kern0.8pt\propline}}
\newdimen\unit
\def\point#1 #2 #3{\rlap{\kern#1\unit
  \raise#2\unit\hbox to 0pt{\hss$#3$\hss}}}
\def\cpt#1 #2 {\point #1 #2 \cdot}

\def\ie{\hbox{\it i.e.}}        
\def\eg{\hbox{\it e.g.}}        
\def\etal{\hbox{\it et al.}}
   \def\Tr{\mathop{\rm Tr}}
\def\({[}  \def\){]}
\def\ra{\rightarrow}       
\def\us{\undertext}        
\def\ifmath#1{\relax\ifmmode #1\else $#1$\fi}
\def\half{\ifmath{{\textstyle{1\over 2}}}}
\def\ihalf{\ifmath{{\textstyle{i\over 2}}}}
\def\L{{\cal L}} \def\M{{\cal M}} \def\T{{\cal T}}

\def\gM{\gamma^\mu}       \def\gm{\gamma_\mu}
       
\def\gf{\gamma_5}
\def\vsla{v\!\!\!\slash}
       \def\MeV{\,{\rm MeV}}
\def\GeV{\,{\rm GeV}}     
\def\Vub{V_{ub}}          \def\Vcb{V_{cb}}
\def\prd#1#2(#3){\sl Phys.~Rev. \bf D#1 \rm (#3) #2}
\def\npb#1#2(#3){\sl Nucl.~Phys. \bf B#1 \rm (#3) #2}
\def\plb#1#2(#3){\sl Phys.~Lett. \bf B#1 \rm (#3) #2}
\def\prep#1#2(#3){\sl Phys.~Rep. \bf #1 \rm (#3) #2}
\def\prl#1#2(#3){\sl Phys.~Rev.~Lett. \bf #1 \rm (#3) #2}
\def\mark#1{\hbox{\(#1\)}}


\REF{\Isgu}{N. Isgur and M.B. Wise, \plb{232}{113}(1989);
\plb{237}{527}(1990).}

\REF{\Shur}{E.V. Shuryak, \plb{93}{134}(1980); \npb{198}{83}(1982).}

\REF{\Nuss}{S. Nussinov and W. Wetzel, \prd{36}{130}(1987).}

\REF{\Volo}{
M.B. Voloshin and M.A. Shifman, Yad.\ Fiz.\ {\bf 45} (1987) 463
\(Sov.\ J.\ Nucl.\ Phys.\ {\bf 45} (1987) 292\); {\bf 47} (1988) 801
\({\bf 47} (1988) 511\).}

\REF{\Eich}{E. Eichten and B. Hill, \plb{234}{511}(1990).;
{\bf 243}, 427 (1990).}

\REF{\Mann}{T. Mannel, W. Roberts, and Z. Ryzak, \npb{368}{204}(1992).}

\REF{\Geor}{H. Georgi, \plb{240}{447}(1990).}

\REF{\FGL}{A.F. Falk, B. Grinstein, and M.E. Luke, \npb{357}{185}(1991).}

\REF{\FGGL}{A.F. Falk, H. Georgi, B. Grinstein, and M.B. Wise,
\npb{343}{1}(1990).}

\REF{\Bjor}{J.D.~Bjorken, {\it Proceedings of the 18th SLAC Summer
Institute on Particle Physics}, pp.\ 167, Stanford, California,
July 1990, edited by J.F. Hawthorne (SLAC, Stanford, 1991).}

\REF{\Luke}{M.E. Luke, \plb{252}{447}(1990); The $1/m_b$ corrections
were included in \mark{\NR}.}

\REF{\NR}{M.~Neubert and V.~Rieckert, \npb{383}{97}(1992).}

\REF{\review}{For a comprehensive review see:
M. Neubert, SLAC--PUB--6263 (1993),
to appear in Physics Reports, and references therein.}

\REF{\SVZ}{M. Shifman, A. Vainshtein, and V. Zakharov,
{\sl Nucl.~Phys.~}{\bf B147} (1979) 385; 448; 519.}

\REF{\RRY}{L. Reinders, H. Rubinstein, and S. Yazaki,
\prep{127}{1}(1985).}

\REF{\SPT}{{\it Vacuum Structure and QCD Sum Rules},
edited by M. Shifman (North Holland, 1992);
P.~Pascual and R.~Tarrach, {\it QCD: Renormalization for the
Practitioner}, Lecture Notes in Physics No.\ 194, edited by H. Araki
\etal\ (Springer, Berlin--Heidelberg--New York--Tokyo, 1984).}

\REF{\SRmn}{M.~Neubert, \prd{45}{2451}(1992).}

\REF{\SRdec}{M. Neubert, \prd{46}{1076}(1992); \nextline
E. Bagan, P. Ball, V.M. Braun, and H.G. Dosch, \plb{278}{457}(1992);
\nextline
D.J. Broadhurst and A.G. Grozin, \plb{274}{421}(1992); \nextline
For a recent review including results not only from HQET, see \eg:
\nextline  P. Colangelo, G. Nardulli, and N. Paver, BARI~TH/93-132.}

\REF{\SRbs}{B.~Blok and M.~Shifman, \prd{47}{2949}(1993).}

\REF{\SRtl}{M.~Neubert, \prd{47}{4063}(1993).}

\REF{\SRiw}{E. Bagan, P. Ball, P. Gosdzinsky, \plb{301}{249}(1993);
\nextline  A.V. Radyushkin, \plb{271}{218}(1991).}

\REF{\SRsubl}{M. Neubert, \prd{46}{3914}(1992).}

\REF{\SRchi}{M.~Neubert, Z.~Ligeti, and Y.~Nir, \plb{301}{101}(1993);
\prd{47}{5060}(1993).}

\REF{\SRxi}{Z. Ligeti, Y. Nir, and M. Neubert, SLAC--PUB--6146,
hep-ph/9305304, to appear in {\sl Phys. Rev.} {\bf D}.}

\REF{\IWbaryon}{N.~Isgur and M.B.~Wise, \npb{348}{276}(1991); \nextline
H.~Georgi, \npb{348}{293}(1991).}

\REF{\baryonsr}{A.G. Grozin and O.I. Yakovlev, \plb{285}{254}(1991);
{\bf B291} (1992) 441.}

\REF{\lattice}{See \eg: E.J.~Eichten, C.T.~Sachrajda,
in this proceedings.}

\REF{\LuMa}{M. Luke and A.V. Manohar, \plb{286}{348}(1992).}

\REF{\FlIs}{J.M. Flynn and N. Isgur, hep-ph/9207223.}

\REF{\FNL}{A.F. Falk, M. Neubert, and M. Luke, \npb{388}{363}(1992).}

\REF{\vcb}{M.~Neubert, \plb{264}{455}(1991).}

\REF{\MAB}{M. Neubert, \prd{46}{2212}(1992); see also:\nextline
A.F. Falk and B. Grinstein, \plb{247}{406}(1990).}

\REF{\FaNe}{A.F. Falk and M. Neubert, \prd{47}{2965}(1993).}

\REF{\Voloshin}{M.B. Voloshin, \prd{46}{3062}(1992).}

\REF{\dRT}{E. de Rafael and J. Taron, hep-ph/9306214,
and references therein.}

\REF{\Adam}{See \eg: A.F.~Falk, in this proceedings,
and references therein.}

\REF{\CLEO}{CLEO collaboration, presented at the 1993 Lepton--Photon
conference.}

\REF{\disc}{I thank M. Neubert for discussions on this point.}

\REF{\BSW}{M. Wirbel, B. Stech, and M. Bauer, {\sl Z. Phys.} {\bf C29}
(1985) 637.}

\REF{\ISGW}{N. Isgur, D. Scora, B. Grinstein, and M.B. Wise,
\prd{39}{799}(1989).}

\REF{\KS}{J.G. Korner and G.A. Schuler,
{\sl Z. Phys.} {\bf C38} (1988) 511. \(E:~{\bf C41} (1989) 690\);
{\bf C46}~(1990)~93.}

\REF{\PBall}{P. Ball, \plb{281}{133}(1992).}

\REF{\vub}{N. Isgur and M. B. Wise, \prd{42}{2388}(1990),
{\bf D41}~(1990)~151.}

\REF{\GZMY}{G. Burdman, Z. Ligeti, M. Neubert, and Y. Nir,
SLAC--PUB--6345, hep-ph/9309272, submitted to {\sl Phys. Rev.} {\bf D}.,
and references therein.}

\REF{\Bigi}{See \eg: I.~Bigi, in this proceedings,
and references therein.}


\Pubnum={\cr\cr WIS-93/103/Oct-PH}
\date={October, 1993}
\titlepage
\title{THE DETERMINATION OF $|\Vcb|$ \break
       AND QCD SUM RULES IN HQET
       \foot{Invited talk at the Advanced Study Conference
       on Heavy Flavours; September 3--7, 1993 at Pavia, Italy.}}
\author{Zoltan Ligeti}
\address{Department of Physics     \break
         Weizmann Institute of Science \break
         Rehovot 76100, Israel}
\vfill
   \centerline{\fourteenrm Abstract}\vskip\headskip
{\leftskip 1truecm \rightskip 1truecm \noindent
I review recent developments in Heavy Quark Effective Theory (HQET)
that lead to an almost model--independent determination of the $|\Vcb|$
element of the Cabibbo--Kobayashi--Maskawa matrix from exclusive
semileptonic $B\ra D^{(*)}$ decays.
In particular, I compare the theoretical uncertainties in the
$B\ra D^*\ell\,\bar\nu$ and the $B\ra D\,\ell\,\bar\nu$ decay modes.
I discuss the applications of QCD sum rules within HQET to semileptonic
heavy meson decays and give predictions for the form factors measurable
in $B\ra D^{(*)}\ell\,\bar\nu$ decays. \par}
\vfill
\endpage


\baselineskip=18pt
\sequentialequations
\chapter{Introduction}

In the absence of direct observations of new physics, heavy hadron
decays may provide the first clues to physics beyond the Standard Model
(SM). They probe the flavor sector, which contains the majority of the
parameters of the SM. These parameters have to satisfy certain relations
provided by, \eg, the unitarity of the Cabibbo--Kobayashi--Maskawa (CKM)
matrix, and the SM prescription of $CP$ violation.
Measuring the parameters of the SM more accurately, one hopes to find
inconsistencies among them, which would give hints to new physics.
Even if the numerical values of these parameters are consistent
with each other, they may provide insights to new physics that can
yield relations among them. (For example, schemes for quark mass
matrices can be tested, that may teach us about horizontal symmetries
or grand unification.) Certain rare decays are particularly sensitive
probes of various extensions of the SM.
However, the theoretical predictions for most of these measurements
are contaminated by large hadronic uncertainties. Reducing these
uncertainties would provide better chances of discovering physics
beyond the SM, and make the bounds on such new physics more
restrictive. Recent developments in this direction constitute
the subject of this~talk.

In hadrons composed of one heavy quark and a number of light degrees
of freedom (gluons and light quarks), the energy scale of strong
interactions is small compared to the heavy quark mass. The heavy quark
acts effectively as a static source of color, resulting in new
symmetries of QCD \mark{\Isgu--\Geor}: the interaction between the
heavy quark and the surrounding light degrees of freedom become
independent of the mass and spin of the heavy quark.
In the $m_Q\gg\Lambda_{\rm QCD}$ limit the velocity of the heavy quark
is conserved with respect to soft processes \mark{\Isgu,\Geor}, and
the complexity of hadronic dynamics results from the strong interactions
among the light degrees of freedom only.
{}From the phenomenological point of view, such a symmetry (even if
broken as the `heavy' quarks are, after all, not infinitely heavy)
is an extremely useful tool, as it provides exact predictions in the
symmetry limit, reducing hadronic uncertainties and model--dependence,
which can only enter in corrections, suppressed by powers of $1/m_Q$.

The heavy quark effective theory (HQET) \mark{\Eich--\Luke,\review}
provides a convenient framework to analyze heavy hadron decays.
It allows for a systematic expansion of hadronic quantities in powers
of $\Lambda_{\rm QCD}/m_Q$ in such a way that the coefficients are
heavy quark spin-- and mass--independent
universal functions of the kinematic variable $y=v\!\cdot\!v'$, where
$v$ and $v'$ denote the velocities of the initial and final hadrons.
These universal functions originate from long distance QCD, so they
can only be investigated using nonperturbative methods. Such a method
is provided by QCD sum rules \mark{\SVZ--\SPT}, which have been widely
used recently to calculate hadronic matrix elements in HQET
\mark{\SRmn--\SRxi}.

In this talk I shall focus on decays of heavy mesons (rather than
baryons). Experimentally they are easier to measure, resulting in
more phenomenological applications (\eg, extraction of $|\Vcb|$),
and most calculations of HQET form factors using either QCD sum rules
or other models have been carried out for these decays.%
\foot{The interested reader can find the HQET formalism for heavy
baryon decays in Ref.~\mark{\IWbaryon}; the QCD sum rules
determination of the relevant universal function in \mark{\baryonsr}.}
Moreover, in heavy meson decays a comparison with lattice calculations
is also possible \mark{\lattice}.

In Section 2, I review aspects of HQET that are relevant to the
discussion in Section 3 of the almost model--independent determination
of $|\Vcb|$ from exclusive semileptonic $B$ meson decays. In Section 4,
I discuss applications of QCD sum rules for heavy meson decays. Besides
the universal functions, predictions for heavy meson decay constants and
semileptonic form factors are also presented. Finally, I summarize and
outline some directions of ongoing (and future) developments.

\chapter{HQET}

The construction of HQET starts with removing the mass--dependent
piece of the momentum operator by introducing a field $h_Q(v,x)$,
which annihilates a heavy quark with velocity~$v$~\mark{\Geor},
$$
h_Q(v,x)=e^{i m_Q v\cdot x}\,P_+(v)\,Q(x) ,\eqno\eq$$
where $P_+(v)={1\over 2}(1+\rlap/v)$ is an on--shell projection
operator onto the heavy quark (rather than antiquark) components
of the spinor, and $Q(x)$ denotes the conventional quark field in QCD.
If $P^\mu$ is the total momentum of the heavy quark, the new field
$h_Q$ carries the residual momentum
$k^\mu = P^\mu -m_Q\,v^\mu \sim\O(\Lambda_{\rm QCD})$,
which does not grow with the heavy quark mass.

In the limit $m_Q\gg\Lambda_{\rm QCD}$, the effective Lagrangian
for the strong interactions of the heavy quark is
\mark{\Eich--\FGL}
$$
\L_{\rm HQET} = \bar h_v\,i v\!\cdot\!D\,h_v
+ {1\over 2 m_Q}\,\Big\( O_{\rm kin}
+ C_{\rm mag}(\mu)\,O_{\rm mag} \Big\) + \O(1/m_Q^2) \,, \eqn\Lag$$
where $D^\mu = \partial^\mu - i g_s T_a A_a^\mu$ is the gauge--covariant
derivative. The leading term respects both the spin and flavor
symmetries. The operators appearing at order $1/m_Q$ are
$$
O_{\rm kin} = \bar h_v\,(i D)^2 h_v \,, \qquad
O_{\rm mag} = {g_s\over 2}\,\bar h_v\,\sigma_{\mu\nu} G^{\mu\nu} h_v \,.
\eqno\eq$$
Here $G^{\mu\nu}$ is the gluon field strength tensor defined by
$\(iD^\mu,iD^\nu\) = i g_s G^{\mu\nu}$.
In the hadron's rest frame, $O_{\rm kin}$ describes the kinetic energy
resulting from the residual motion of the heavy quark, whereas
$O_{\rm mag}$ corresponds to the chromomagnetic coupling of the heavy
quark spin to the gluon field. While $O_{\rm kin}$ violates only the
heavy quark flavor symmetry, $O_{\rm mag}$ violates the spin symmetry as
well. Due to reparameterization invariance \mark{\LuMa} $O_{\rm kin}$
is not renormalized to all orders in perturbation theory, while
$C_{\rm mag}(\mu)$ is a renormalization factor for $O_{\rm mag}$.
The heavy quark symmetries are also manifest in the Feynman rules of
the effective theory: the propagator of a heavy quark becomes
independent of its mass (flavor symmetry) and no gamma matrix appears
in the coupling of a heavy quark to the gluon field (spin symmetry).
\endpage
\baselineskip=12pt
$$
\matrix{
\prop & \quad \longrightarrow \quad & \heavyprop\cr
\noalign{\vskip7pt}
        \displaystyle {i\over p\!\!\!\!\;\slash -m_Q} && \displaystyle
        {1+\vsla\over2}\,{i\over v\!\cdot\! k} \cr}$$
\centerline{(a) \hbox{\hskip 8truecm} (b)}
\newbox\cycloid
\setbox\cycloid=\hbox{\unit=0.5pt
\cpt 0.0000 6.0000 \cpt 0.4998 5.9896 \cpt 0.9986 5.9585 \cpt
1.4954 5.9067 \cpt 1.9895 5.8343 \cpt 2.4799 5.7412 \cpt 2.9658
5.6277 \cpt 3.4463 5.4938 \cpt 3.9205 5.3396 \cpt 4.3874 5.1653
\cpt 4.8461 4.9709 \cpt 5.2957 4.7567 \cpt 5.7351 4.5228 \cpt
6.1633 4.2695 \cpt 6.5792 3.9969 \cpt 6.9819 3.7055 \cpt 7.3700
3.3954 \cpt 7.7424 3.0671 \cpt 8.0979 2.7209 \cpt 8.4351 2.3573
\cpt 8.7526 1.9769 \cpt 9.0489 1.5802 \cpt 9.3225 1.1679 \cpt
9.5716 0.7409 \cpt 9.7944 0.3001 \cpt 9.9889 -0.1533 \cpt 10.1530
-0.6180 \cpt 10.2844 -1.0924 \cpt 10.3805 -1.5744 \cpt 10.4386
-2.0619 \cpt 10.4555 -2.5516 \cpt 10.4281 -3.0401 \cpt 10.3528
-3.5226 \cpt 10.2257 -3.9932 \cpt 10.0433 -4.4442 \cpt 9.8020
-4.8659 \cpt 9.4994 -5.2457 \cpt 9.1354 -5.5677 \cpt 8.7141
-5.8131 \cpt 8.2464 -5.9621 \cpt 7.7516 -5.9974 \cpt 7.2570
-5.9114 \cpt 6.7910 -5.7096 \cpt 6.3766 -5.4093 \cpt 6.0267
-5.0329 \cpt 5.7457 -4.6020 \cpt 5.5322 -4.1342 \cpt 5.3823
-3.6432 \cpt 5.2912 -3.1389 \cpt 5.2537 -2.6287 \cpt 5.2653
-2.1183 \cpt 5.3216 -1.6118 \cpt 5.4188 -1.1123 \cpt 5.5536
-0.6224 \cpt 5.7227 -0.1439 \cpt 5.9236 0.3214 \cpt 6.1536 0.7724
\cpt 6.4107 1.2080 \cpt 6.6926 1.6274 \cpt 6.9975 2.0298 \cpt
7.3238 2.4147 \cpt 7.6697 2.7814 \cpt 8.0338 3.1297 \cpt 8.4146
3.4591 \cpt 8.8108 3.7693 \cpt 9.2213 4.0600 \cpt 9.6446 4.3310
\cpt 10.0798 4.5820 \cpt 10.5257 4.8130 \cpt 10.9813 5.0238 \cpt
11.4455 5.2142 \cpt 11.9174 5.3841 \cpt 12.3960 5.5335 \cpt
12.8803 5.6622 \cpt 13.3694 5.7702 \cpt 13.8625 5.8576 \cpt
14.3587 5.9241 \cpt 14.8569 5.9699 \cpt 15.3565 5.9949 }
\def\gluon{\vcenter{%
 \hbox{\copy\cycloid}}\hbox to 7.85pt{\hfil}}
$$
\matrix{
\gluon\gluon\gluon\gluon \vcenter{\hrule height 40pt width1pt}
 & \quad\longrightarrow\quad &
\gluon\gluon\gluon\gluon \vcenter{\hbox{\vrule height 40pt
width1pt \kern0.8pt \vrule height40pt width1pt}} \cr
\noalign{\vskip7pt}
\displaystyle ig \gM\, {\lambda^a\over2} & & \displaystyle
              ig v^\mu\, {\lambda^a\over2} \cr}
$$
\setbox\cycloid=\hbox{}
\smallskip
\baselineskip=18pt
\cap{\tenpoint {\bf Figure 1:}
Feynman rules in ``full" QCD (a) and in HQET (b). Double lines denote
heavy quark propagators in HQET (from \mark{\FlIs}).}
\medskip

Any operator of the full theory that contains one or more heavy quark
fields can be matched onto a short distance expansion in terms of
operators of the effective theory. In particular, the expansion of the
heavy quark current $\bar Q'\,\Gamma\, Q$ that gives rise to
$B\ra D^{(*)}$ decays is matched onto
$$
\bar Q'\,\Gamma\,Q\rightarrow
\sum_{i} C_i(\mu)\,J_i
+\sum_j\left\({B_j(\mu)\over2m_Q}+{B'_j(\mu)\over2m_{Q'}}\right\)O_j
+\O(1/m_Q^2) \,.\eqn\current$$
The operators $\{J_i\}$ form a complete set of local dimension three
current operators with the same quantum numbers as the current in the
full theory. There are three such operators
$J_i=\bar h_{Q'}\,\Gamma_i\,h_Q$, with
$\Gamma_i=\{\gM,\,v^\mu,\,{v'}^\mu\}$ for the vector current, and
$\Gamma_i=\{\gM\gf,\,v^\mu\gf,\,{v'}^\mu\gf\}$ for the axial current.
(In the leading logarithmic approximation only $\Gamma=\gM\,(\gf)$
contributes. Radiative corrections induce the other operators.)
Similarly, $\{O_j\}$ denote a complete set of local dimension four
operators. Since there are fourteen independent such operators, we
do not display them explicitly here.
These effective current operators have non--zero anomalous dimensions.
The coefficients $C_i(\mu)$ and $B_j$\hbox{${}^({}'{}^)$}$(\mu)$
ensure that the final result for any physical quantity is independent
of the renormalization procedure. At present, the expansion of the
effective Lagrangian and the weak currents are known in perturbation
theory up to order $\alpha_s/m_Q$ and $1/m_Q^2$ \mark{\review}.

\section{Leading order}

Matrix elements in HQET are conveniently calculated in the compact trace
formalism \mark{\FGGL,\Bjor}, where a heavy meson is represented by its
spin wave function
$$\M(v)=\sqrt{m_Q}\,{(1+\rlap/v)\over2}\cases{
-\gamma_5&;\ $J^P=0^-$,\cr
\rlap/\epsilon&;\ $J^P=1^-$,\cr}\eqno\eq$$
which has the correct transformation properties under Lorentz boosts
and heavy quark spin rotations. When the external weak current changes
$v\ra v'$ (and maybe $Q\ra Q'$), the light degrees of freedom have to
rearrange themselves, which yields a form factor suppression.
Due to heavy quark symmetry, this suppression factor cannot depend on
the spin and the mass of the heavy quark, neither on the Dirac structure
of the current. Lorentz and parity invariance, and the
properties of $\M(v)$ imply that dependence only on $y=v\!\cdot\! v'$
and on the renormalization scale $\mu$ is allowed.
Hence, a single universal, \ie\ only $y$ and $\mu$ dependent, function
$\xi(y,\mu)$ is sufficient to parameterize all semileptonic
$M(v)\ra M'(v')\,\ell\,\bar\nu$ decays, where $M$ and $M'$ are pseudoscalar
or vector mesons containing a single heavy quark:
$$
\bra{M'(v')}\bar h'_{v'}\,\Gamma\,h_v\ket{M(v)}
= -\xi(y,\mu)\,\Tr\big\{\,\overline{\M}'(v')\,\Gamma\,{\M}(v)\big\}\,.
\eqno\eq$$
Vector current conservation implies that when the heavy meson in the
final state is at rest in the rest frame of the decaying heavy meson,
this so--called Isgur--Wise function satisfies \hbox{$\xi(1)=1$}.
The predictions of HQS are most restrictive at this special kinematic
point (``zero recoil": $y=1$), allowing model--independent predictions,
unaffected by hadronic uncertainties.

Thus, at leading order in the heavy quark expansion, matrix elements
factorize into a kinematic part that depends on the mass and the
spin--parity of the mesons, and a reduced matrix element that describes
the light degrees of freedom.
This is a remarkable simplification, as {\it a--priori} six independent
form factors describe the semileptonic $B\to D^{(\ast)}$ transitions.
Since the $b$ and $c$ quarks are not much heavier than
$\Lambda_{\rm QCD}$, an analysis of the $1/m_Q$ corrections is
important for most phenomenological applications.

\section{$1/m_Q$ corrections}

At subleading order, matrix elements receive contributions from the
higher dimension operators in the effective Lagrangian \Lag\ and
in the effective current \current.
The idea is to leave the heavy quark propagator identical to its
leading order expression and account for the correction terms in
the Lagrangian as insertions of operators.
To parameterize their matrix elements we need three new universal
functions $\chi_i(y)$~$(i=1,2,3)$. Vector current conservation implies
$\chi_1(1)=\chi_3(1)=0$ (this is known as Luke's theorem \mark{\Luke}).

Matrix elements of the $1/m_Q$ corrections in the effective current
\current\ are parameterized in terms of another three universal
form factors, usually denoted by $\xi_+(y)$, $\xi_-(y)$, and $\xi_3(y)$.
Imposing the equation of motion, $i(v\!\cdot\! D)h_Q=0$, on
the matrix element yields two constraints~\mark{\Luke}. Thus only
one of these three functions, say $\xi_3(y)$, is independent.
\bigskip
\vbox{\begintable
function \| $\xi(y)$ \| $\chi_1(y)$ | $\chi_2(y)$ | $\chi_3(y)$ |
$\xi_3(y)$ \crthick
{}~normalization~ \| ~$\xi(1)=1$~ \| ~$\chi_1(1)=0$~ | no | $\chi_3(1)=0$ |
no \cr
broken symmetries \| no \| flavor | ~spin, flavor~ | ~spin, flavor~ |
{}~spin, flavor~
\endtable
\smallskip
\centerline{\tenpoint {\bf Table 1:}
Properties of the universal functions of HQET.}
}\smallskip

So already at order $1/m_Q$ one encounters a set of four universal
functions $\xi_3(y)$ and $\chi_i(y)~(i=1,2,3)$ in addition to the
Isgur--Wise function, as well as a parameter $\bar\Lambda=m_M-m_Q$
that describes the mass difference between a heavy meson and the
heavy quark that it contains \mark{\Luke,\FNL}. This parameter sets the
scale of the $1/m_Q$ expansion; in fact, the real expansion parameter
is $\bar\Lambda/2m_Q$. The universal form factors are real due to
$T$ invariance of the strong interaction. Knowledge of these functions
would teach us about confinement and enhance the phenomenological
applications of the heavy quark expansion.

\chapter{Model Independent Determination of $|\Vcb|$}

The magnitude of the CKM matrix element $\Vcb$ can best be determined
from an extrapolation of the semileptonic $B$ decay rate to zero
recoil, making use of the known normalization of the Isgur--Wise
function at that point \mark{\vcb}. The advantage of this method over
previous determinations of $|\Vcb|$ is that in the framework of HQET
a clear separation between the model--independent and model--dependent
ingredients of the analysis is possible, due to a systematic expansion
in the small parameters $\bar\Lambda/2m_{c,b}\,$.

The $B\ra D^{(*)}\ell\,\bar\nu$ differential decay rate near zero recoil
is given by,
$$\eqalign{
\lim_{y\ra1}{d\Gamma(B\ra D^*\ell\,\bar\nu)\over dy}
=\,& {G_F^2\over4\pi^3}\,
m_{D^*}^3(m_B-m_{D^*})^2\, |\Vcb|^2\, (y^2-1)^{1/2}\, \eta^{*2},\crr{4pt}
\lim_{y\ra1}{d\Gamma(B\ra D\,\ell\,\bar\nu)\over dy}
=\,& {G_F^2\over 48\pi^3}\,
m_D^3(m_B+m_D)^2\, |\Vcb|^2\, (y^2-1)^{3/2}\, \eta^2.\cr}\eqn\rate$$
On the right hand sides of these relations the Fermi constant and the
meson masses are well known quantities, powers of $(y^2-1)$ arise
from phase space, $\eta^{(*)}$ is defined to include all hadronic
uncertainties at $y=1$, and we want to extract $|\Vcb|$. The kinematic
variable $y$ is related to the conventional $q^2$ via
$$
y \equiv v\cdot v' ={m_B^2+m_{D^{(*)}}^2-q^2\over 2\,m_B\,m_{D^{(*)}}}\,.
\eqno\eq$$
In this notation maximal $q^2$ corresponds to $y=1$, while $q^2=0$
corresponds to maximal $y$, which is about 1.5 and 1.6 in $B\ra D^*$
and $B\ra D$ decays respectively. The problem is that {\it a--priori} we
know nothing about $\eta^{(*)}$, except that it should be of order one.
The power of HQS is that it gives the model--independent prediction in
the infinite quark mass limit: $\lim_{m_Q\ra\infty}\eta^{(*)}=\xi(1)=1$.
This allows us to write
$$
\eta^{(*)}=1 +\delta_{\alpha_s}^{(*)} +\delta_{1/m_Q}^{(*)}
+\delta_{1/m_Q^2}^{(*)} +{\rm higher~order}\,.\eqno\eq$$
We shall discuss each of the correction terms in the sequel.
The calculation of the perturbative QCD corrections must include the
full order $\alpha_s$ terms (not just the leading logarithms), because
$\ln(m_b/m_c)\sim1.2$ is not a big number. We emphasize that these
corrections do not introduce uncertainty into the analysis.
They are given by $\delta_{\alpha_s}=0.05$ and
$\delta_{\alpha_s}^*=-0.01$ \mark{\MAB}.
While the $B\ra D^*\ell\,\bar\nu$ decay rate is protected against
$1/m_Q$ corrections at zero recoil due to Luke's theorem \mark{\Luke},
\ie\ $\delta_{1/m_Q}^*=0$, the $B\ra D\,\ell\,\bar\nu$ decay is not,
due to its helicity suppression \mark{\NR,\vcb}.
In this latter case the $1/m_Q$ corrections are given by
$$
\delta_{1/m_Q} =
\left({\bar\Lambda\over2m_c}+{\bar\Lambda\over2m_b}\right)
\left({m_B-m_D\over m_B+m_D}\right)^2 \(1-2\xi_3(1)\)\,.\eqn\corr$$
Clearly, the form factor $\xi_3(y)$ is very important for the
determination of $|V_{cb}|$ from $B\ra D\,\ell\,\bar\nu$ decays.
For example, if $\xi_3(1)$ were around $-1$ then this $1/m_Q$
correction would be about 15\%, while if it were around $0.5$
then the $1/m_Q$ correction would vanish.
In the next section we shall see that QCD sum rules predict
$\xi_3(1)=0.6\pm0.2$ \mark{\SRxi}, which implies that the $1/m_Q$
correction to the $B\ra D\,\ell\,\bar\nu$ decay rate at zero recoil
is not more than 3\%. This is a result of two suppression factors
(beyond $\bar\Lambda/2m_Q$): the Voloshin--Shifman factor
$\((m_B-m_D)/(m_B+m_D)\)^2\simeq0.23$ \mark{\Volo},
and an accidental suppression factor of $\(1-2\xi_3(1)\)\sim0.2$.
The $1/m_Q^2$ corrections are expected to be 3--4\% on dimensional
grounds, and the detailed analysis of Ref.~\mark{\FaNe} supports
that these corrections are not larger than the above estimate.
(This statement, however, is somewhat model--dependent, which leaves
room for arguments that these corrections might be larger.)
The higher order corrections are certainly negligible, \eg\ the
characteristic size of the second order QCD corrections is
$\(\alpha_s(m_c)/\pi\)^2<1$\%.
Thus the $1/m_Q$ correction to the $B\ra D\,\ell\,\bar\nu$ decay rate
at zero recoil is not more than the expected $1/m_Q^2$ corrections.
This suggests that the theoretical uncertainty in the determination
of $|V_{cb}|$ from $B\ra D$ transition is comparable to that in
$B\ra D^*$, even though the latter appears only at order $1/m_Q^2$.
Of course, the experimental measurement of $B\ra D\,\ell\,\bar\nu$ near
zero recoil is more difficult due to extra power of $(y-1)$ helicity
suppression in Eq.~\rate. The reward of such a measurement, however,
would be an independent determination of $|\Vcb|$ with surprisingly
small theoretical uncertainties.

A different kind of uncertainty enters into the analysis because
phase space vanishes at $y=1$. Therefore, an extrapolation
of the measured spectrum to zero recoil is needed to obtain the
numerical value of $|\Vcb|$. At present this gives rise to both
a theoretical and an experimental error. The former is due to the
fact that the precise shape of the Isgur--Wise function is not known,
while the latter is dominated by statistical error.
However, in the not--so--distant future this theoretical uncertainty
will almost disappear, since in an asymmetric $B$ factory the zero
recoil limit does not correspond to the $D^*$ meson being at rest in
the laboratory frame. Then the pion in the subsequent $D^*\ra D\,\pi$
decay is boosted, while it is almost at rest for ARGUS and CLEO.
In addition, the $(y^2-1)^{1/2}$ phase space suppression is a very
mild one: the statistical error of measuring the rate at $y=1.05$ is
less than a factor of two higher than that at the endpoint
$y_{\rm max}\simeq 1.5$.
At present any information is important on the shape of the Isgur--Wise
function, in particular on its slope at $y=1$, to make the extraction of
$|\Vcb|$ more reliable. Global quark--hadron duality (that the sum of
probabilities to decay into hadrons equals to the probability of free
quark transition when $m_Q\ra\infty$) gives the Bjorken \mark{\Bjor}
and Voloshin \mark{\Voloshin} sum rules:
$$
{1\over4} < \rho^2 \lsim 1 \,,\eqno\eq$$
where $\rho^2$ is minus the slope of the Isgur--Wise function at $y=1$.
For a discussion of the bounds derived by de Rafael and Taron \mark{\dRT}
we refer to \mark{\Adam}.

In summary, we emphasize that due to heavy quark symmetry the
theoretical prediction for the $B\ra D^{(*)}\ell\,\bar\nu$ decay rate
near zero recoil is very precise:
$$\eqalign{
\eta^* =\,& 1+\delta_{\alpha_s}^* +\delta_{1/m_Q^2}^* =0.99\pm0.05\,,\cr
\eta =\,& 1+\delta_{\alpha_s} +\delta_{1/m_Q} +\delta_{1/m_Q^2}
= 1.05\pm0.08 \,.\cr}\eqno\eq$$
Experimentally only the $B\ra D^*\ell\,\bar\nu$ spectrum has been
measured (for $B\ra D\,\ell\,\bar\nu$ only the total decay rate is
known). The following figure shows the most recent data from the
CLEO collaboration \mark{\CLEO}.
\midinsert
\smallskip
\epsfysize=7.8truecm
\centerline{\epsffile{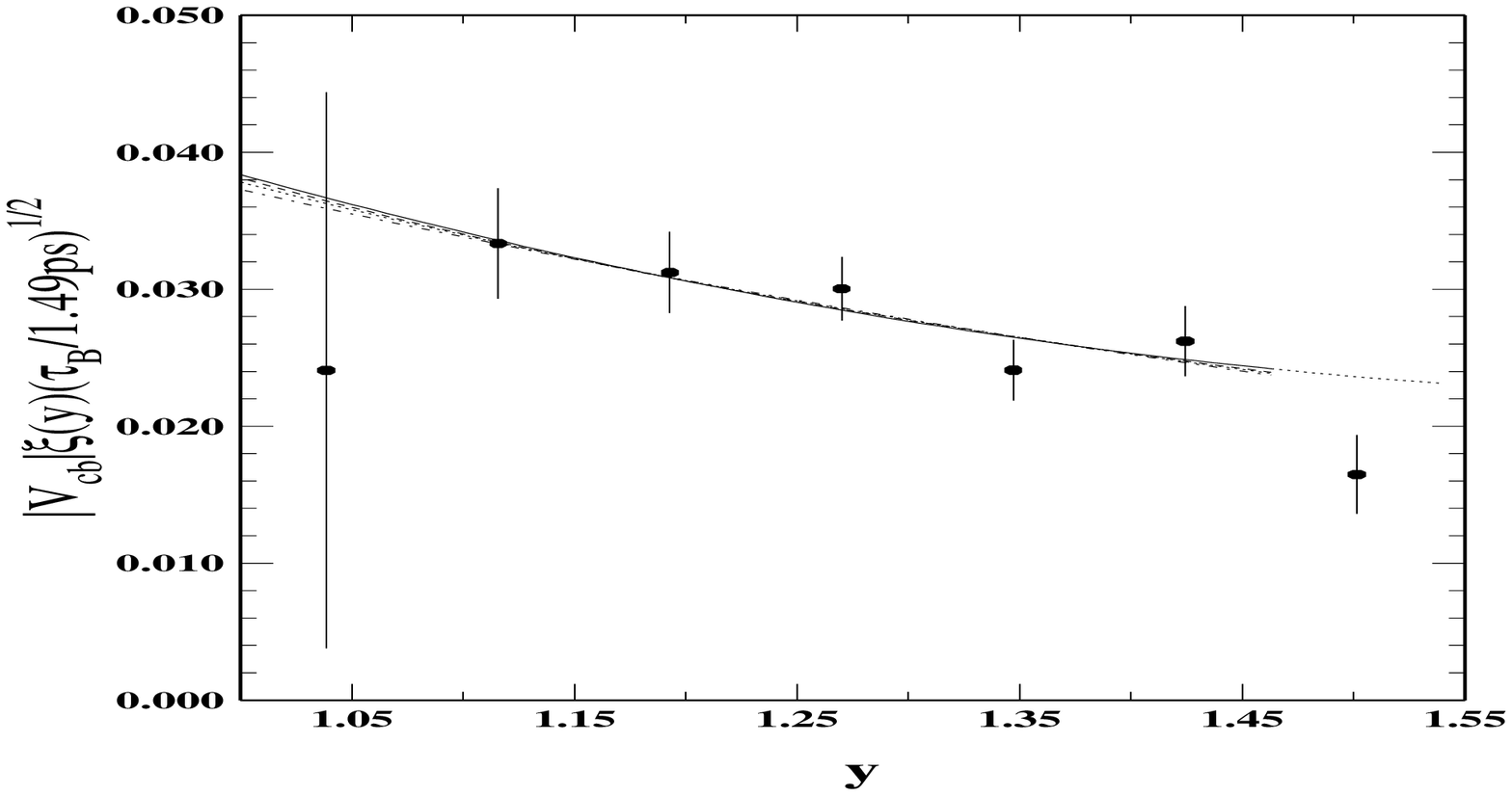}}
\smallskip
\cap{\tenpoint {\bf Figure 2:}
The measured $d\Gamma(B\ra D^*\ell\,\bar\nu)/dy$ distribution.
The curves represent fits with different functional forms for the
Isgur--Wise function.}
\smallskip
\endinsert
\noindent
They obtain from a fit to this distribution
$$\eqalign{
|\Vcb| =\,& 0.037\pm0.005\pm0.004 \,,\cr
\rho^2 =\,& 1.0\pm0.4\pm0.2 \,.\cr}\eqno\eq$$
As the theoretical uncertainty in the determination of $|\Vcb|$
is significantly smaller than the roughly 15\% systematic plus
statistical error at present -- we should look forward to new data
from CLEO and a future $B$ factory.
\endpage

\chapter{QCD Sum Rules}

We have seen, so far, two important examples when more information
is needed about the universal functions describing heavy meson decays
than what is provided by vector current conservation (see: Table 1).
The shape of the Isgur--Wise function is important for the extraction
of $|\Vcb|$, and knowledge of $\xi_3(1)$ is essential for the
determination of $|\Vcb|$ from the decay $B\ra D\,\ell\,\bar\nu$.
These universal functions originate from the long distance confining
interactions, so they can only be investigated using nonperturbative
approaches to QCD. While short distance QCD is well understood,
there is still no accurate quantitative framework for dealing with
the long distance, strong interaction regime.
Lattice gauge theory is the only method known that comes from first
principles and has the hope of achieving arbitrary accuracy.
This is hindered, however, at present, by the technical limitations
of this approach.
A less fundamental method is provided by QCD sum rules \mark{\SVZ--\SPT},
which is particularly suited for the calculation of HQET form factors.
It has the advantage over other models that it does not rely on
{\it ad hoc} assumptions, the errors can be estimated, and there is a
consistency check on its assumptions within the method.

In the rest of this section we first describe the QCD sum rule method
through the example of calculating the heavy meson decay constant in
HQET, then give the results for the universal functions, and finally
discuss the predictions of these results for the form factors in
semileptonic heavy meson decays.

\section{Simplest example: decay constant}

The following analysis of the two--point function is particularly
important as it determines besides the heavy meson decay constant,
the parameter $\bar\Lambda$ as well.
The analogue of the decay constant $f_M$ in HQET is defined by
$$
\bra{0}\bar q\,\Gamma\,h_v\ket{M(v)} = \ihalf F(\mu)\Tr\{\Gamma\,\M(v)\}
\,.\eqno\eq$$
In leading order $f_M\sqrt{m_M}\simeq F(\mu)$.
The idea of QCD sum rules is to calculate the current--current correlator
in two different ways. In HQET one defines \mark{\SRmn}
$$
\Pi(\omega)=i\int d^4x\,e^{ik\cdot x}
\bra{0}\T\{J_M^\dagger(x),J_M(0)\}\ket{0}\,. \eqn\corr$$
Here $J_M$ is an interpolating current for the ground state heavy mesons
$J_M=\bar h_v\,\Gamma_M\,q$ ($\Gamma_P=-\gf$ and $\Gamma_V=\gm-v_\mu$
for the pseudoscalar and vector mesons respectively), $k=P-m_Q\,v$ is
the ``residual" off--shell momentum \mark{\Geor}, and
$\omega=2v\!\cdot\! k$. It is convenient to factor out the Lorentz
structure of the two--point function, by defining $\pi(\omega)$ through
$\Pi(\omega)=-\half\Tr\{\bar\Gamma_M\, P_+\, \Gamma_M\}\,\pi(\omega)$.

\noindent\us{Theoretical side:}~ One side of the sum rule is a
diagrammatic calculation of the correlator in terms of quark and
gluon fields.
As we approach resonances from short distances, nonperturbative
effects induce power corrections that violate asymptotic freedom.
These effects are taken into account in the Wilson operator
product expansion (OPE)
$$
\pi^{\rm theo}(\omega)=\sum_n C_n O_n
=\pi_{\rm pert}(\omega) + \pi_{\rm cond}(\omega) \eqn\OPE$$
through the so--called ``condensates". These are vacuum expectation
values (VEV-s) of gauge invariant local quark--gluon operators
\mark{\SVZ}:
$$
\bra{0}\bar qq\ket{0},\qquad
\bra{0}\alpha_s G_{\mu\nu}G^{\mu\nu}\ket{0},\qquad
\bra{0}g_s\bar q\sigma_{\mu\nu}G^{\mu\nu}q\ket{0},\qquad\ldots \eqno\eq$$
The perturbative contributions are depicted in the first line of Fig.~3,
while the condensates are shown in the second and third lines (the gluon
condensate $\VEV{\alpha_s GG}$ does not contribute).
\midinsert
\smallskip
\epsfysize=5truecm
\centerline{\epsffile{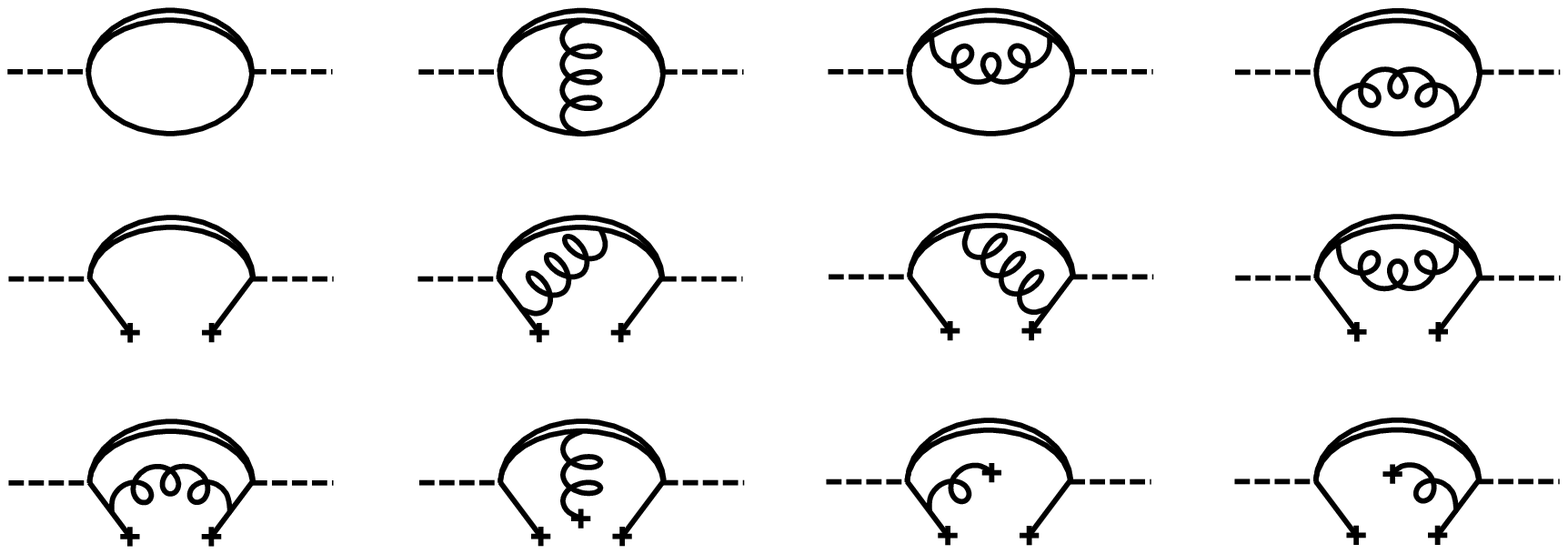}}
\smallskip
\cap{\tenpoint {\bf Figure 3:} Feynman diagrams contributing to the sum
rule for meson decay constants in HQET. The disjoint lines symbolize
the nonvanishing VEV-s of the condensates (from \mark{\review}).}
\smallskip
\endinsert
\noindent
The condensates vanish by definition in the standard perturbation theory.
Assuming that they can acquire non--zero VEV-s is a simple (but maybe
very reasonable) modeling of the QCD vacuum.
Since the condensates have different dimensions, Eq.~\OPE\ forms an
expansion in inverse powers of $\omega$.
The coefficients $C_n$ include, by construction, only short distance
effects, the large distance contributions are accounted for by the
vacuum to vacuum matrix elements.
Strictly speaking, this gives rise to an uncertainty of the method, as
we shall assume that $C_n$ can be calculated perturbatively, while the
nonperturbative physics is hidden solely in the VEV-s of $O_n$.
These parameters are determined from various processes \mark{\SVZ,\RRY},
$$\eqalign{
\VEV{\bar qq} =\,& - 0.013\,{\rm GeV}^3 \,, \cr
\VEV{\alpha_s GG} =\,& 0.04\,{\rm GeV}^4 \,, \cr
\VEV{\bar q\,g_s\sigma_{\alpha\beta}G^{\alpha\beta} q} =\,&
m_0^2\,\VEV{\bar q q} \,,~~~ m_0^2 = 0.8\,{\rm GeV}^2 \,,\cr} \eqno\eq$$
and then their values can be used as inputs in other calculations.
Out of these, the gluon condensate is the most poorly determined,
however, it plays a negligible role in the sum rules we shall discuss.
This value of $\alpha_s\simeq 0.3$ corresponds to the scale
$\mu=2\bar\Lambda\simeq 1\,$GeV, which is appropriate for evaluating
radiative corrections in the effective theory.
Hence we write the correlator as
$\pi^{\rm theo}=\pi_{\rm pert}+\pi_{\rm cond}$, and
$\pi_{\rm pert}$ can further be rewritten as a dispersion integral
$$
\pi_{\rm pert}(\omega)=
\int d\nu\, {\rho_{\rm pert}(\nu)\over \nu-\omega-i\epsilon}
+{\rm subtractions}\,.\eqno\eq$$

\noindent\us{Phenomenological side:}~
The second calculation of the correlator is in terms of hadrons.
$\Pi(\omega)$ is analytic in $\omega$ with discontinuities along
the positive real axis. The correlator can be expressed as a sum
over a complete set of physical intermediate states:
$$
\Pi^{\rm phen}(\omega)=
\sum_X {|\!\bra{X(v)}J_M\ket{0}\!|^2\over \omega_X-\omega-i\epsilon}
+{\rm subtractions}\,,\eqno\eq$$
where the sum is over both discrete and continuum states, and
$\omega_X =2\,(m_X-m_Q)$.
To evaluate the right hand side without a detailed knowledge of
the spectrum and the matrix elements of the excited states, we
employ local (in $\omega$) quark--hadron duality.
We separate the pole corresponding to the ground state meson, and
approximate the contribution of higher resonances by an integral
over the perturbative spectral density above the so--called
continuum threshold $\omega_0$. We find
$$
\pi^{\rm phen}(\omega)= {F^2(\mu)\over2\bar\Lambda-\omega-i\epsilon}
+\int_{\omega_0}^\infty d\nu\,
{\rho_{\rm pert}(\nu)\over\nu-\omega-i\epsilon}
+{\rm subtractions}\,.\eqno\eq$$

The QCD sum rules are obtained by matching the theoretical expression
with the phenomenological one:
$$
{F^2(\mu)\over2\bar\Lambda-\omega-i\epsilon}=
\int_0^{\omega_0} d\nu\, {\rho_{\rm pert}(\nu)\over\nu-\omega-i\epsilon}
+{\rm subtractions}+\pi_{\rm cond}(\omega)\,.\eqn\srr$$
In order to extract information about the ground state, one has to go
to small values of $(-\omega)$ to enhance the relative weight
of the low energy contributions. On the other hand, the theoretical
calculation is only reliable when $(-\omega)$ is large.
To achieve a balance between these contradicting requirements,
an alternative way to probe $\pi(\omega)$ for small values of
$(-\omega)$ is by taking derivatives. In the limit $-\omega\ra\infty$
and the number of derivatives $n\ra\infty$, one is sensitive to the
behavior of the correlator at scales $T=-\omega/n$.
This defines the Borel transformation
$$
{1\over T}\,\hat B_T^{(\omega)}=\lim_{\scriptstyle -\omega,n\ra\infty
\atop \scriptstyle -\omega/n=T}
{\omega^n\over\Gamma(n)}\left(-{d\over d\omega}\right)^n.\eqno\eq$$
Employing this operator alters the sum rule \srr\ in such a way that
\parskip = 0pt
\item{-}
possible subtraction terms are eliminated;
\item{-}
contributions of higher dimensional condensates are factorially
suppressed;
\item{-}
higher resonance contributions are exponentially damped in the
dispersion integral (the weight function $\(\nu-\omega\)^{-1}$ is
replaced by $\exp\(-\nu/T\)$).
\par \noindent \parskip = 4pt plus 1pt minus 1pt
This yields the final form of the QCD sum rule:
$$
F^2(\mu)\,e^{-2\bar\Lambda/T}= \int_0^{\omega_0}
d\nu\,\rho_{\rm pert}(\nu)\,e^{-\nu/T}
+\hat B_T^{(\omega)}\,\pi_{\rm cond}(\omega).\eqn\sr$$

To evaluate this sum rule, the continuum threshold $\omega_0$,
the Borel parameter $T$, and $\bar\Lambda$ have to be determined.
Notice that by taking the logarithmic derivative of Eq.~\sr\
with respect to $T^{-1}$ we get a sum rule for $\bar\Lambda$.
Next, we have to find the continuum threshold $\omega_0$ such that
$\bar\Lambda$ is stable with respect to variations of the Borel
parameter $T$. Then we use this value of $\bar\Lambda$ to extract
$F^2(\mu)$ from the sum rule \sr. The range of $\omega_0$ and $T$
in which the predictions of the sum rule (for $\bar\Lambda$ and
$F^2(\mu)$ in the present case) are independent of the value of
$T$ is called the ``sum rule window".
This provides a consistency check on the assumption of local duality,
in the sense that if duality did not hold there were no reason why
such a stability region existed (it can only result from
cancelation of contributions on the theoretical against the
phenomenological side of the sum rule).
For the consistency of the calculation it is necessary that both sum
rules be stable in the same range of $\omega_0$ and $T$. Whether
this happens is far from trivial and checks, in fact, the
consistency of the approximations and the applicability of the method.
Another assumption of QCD sum rules is that there exists a transition
domain where the perturbative calculation is still reliable
\($\alpha_s(T)$ is not too large\) and the spectral density is
sufficiently sensitive to the ground state.
This can be assured by requiring that in the sum rule window the
nonperturbative contributions to the sum rule be less than 30\% of the
perturbative ones, and that the pole contribution accounts for at least
30\% of the perturbative part of the correlator.

Several authors calculated the decay constants of heavy mesons
\mark{\SRmn,\SRdec}. The symmetry breaking corrections (both from QCD
and $1/m_Q$ effects) are large, resulting in significant deviations
from the $f_B/f_D=\sqrt{M_D/M_B}$ scaling of the symmetry limit.
We quote \mark{\SRmn}
$$
f_D=170\pm30\MeV\,, \qquad\qquad f_B=190\pm50\MeV\,,\eqno\eq$$
which compares well with lattice results \mark{\lattice}.
The mass parameter $\bar\Lambda$ that sets the characteristic scale of
the $1/m_Q$ expansion is significantly higher than $\Lambda_{\rm QCD}$:
$\bar\Lambda = 0.57\pm0.07\GeV$ \mark{\SRmn,\review}.

\section{Universal functions}

For the calculation of the universal form factors, we need to consider
three--point correlation functions. The calculation, in principle,
is very similar to the one outlined above.
There are certain technical complications related to the applicability
of local duality in two variables; different choices of the duality
region to model the higher resonance states; and the possibility of
introducing non--local condensates. These have all been discussed
in detail in the literature \mark{\SRmn,\SRbs}.
The important point is that the physical predictions are obtained
by taking the ratio of the three--point and the two--point sum rules.
This assures that the normalization conditions at zero recoil
are exactly fulfilled, independent of the value of the Borel parameter
$T$ and the continuum threshold $\omega_0$.
These sum rules are probably more accurate than the one for the decay
constant, since some of the systematic errors drop out from such a
ratio, and there is no dependence on the less accurately determined
$\bar\Lambda$ parameter.
It is, in fact, very promising that the position of the sum rule
window is almost identical in the calculations of the decay constant,
Isgur--Wise function, and the four subleading universal functions
at order $1/m_Q$ ($\omega_0=2.0\pm0.3\GeV$, $T=0.8\pm0.2\GeV$).
Since the order $\alpha_s$ terms are large in the two--point function,
to obtain a reliable result, it is imperative to include the order
$\alpha_s$ corrections into the analysis of the three--point functions
as well. The evaluation of the arising two loop diagrams require
somewhat sophisticated calculational techniques \mark{\SRchi,\SRtl}.
By now, all parameters that appear in the heavy quark expansion of
meson form factors have been calculated to order $\alpha_s/m_Q$ in
the framework of QCD sum rules.%
\foot{This is not true for $\chi_1(y)$. However, this function does not
affect any of the model--independent predictions of HQET.}

\subsection{Isgur--Wise function:}~%
Several authors calculated the Isgur--Wise function
\mark{\SRmn,\SRbs--\SRiw} (see: Fig.~4).
Of special interest is its slope at zero recoil, described by the
$\rho^2$ parameter. It turns out, however, that the sum rule prediction
for this quantity is sensitive to the choice of the duality region,
in particular to whether the continuum threshold $\omega_0$
is allowed to depend on $y$ \mark{\disc}.
Combining these uncertainties we conclude that QCD sum rules have
a limited power in determining $\rho^2$, only constraining it to
$0.6 < \rho^2 \lsim 1.2$ \mark{\SRmn,\SRbs}.
This upper limit can only be realized by allowing $\omega_0$ to depend
on $y$, and it slightly exceeds the Voloshin bound $\rho^2\lsim1$.
In fact, the experimental data seem to prefer $\rho^2\approx 1$
\mark{\CLEO}.
\midinsert
\epsfysize=8truecm
\centerline{\epsffile{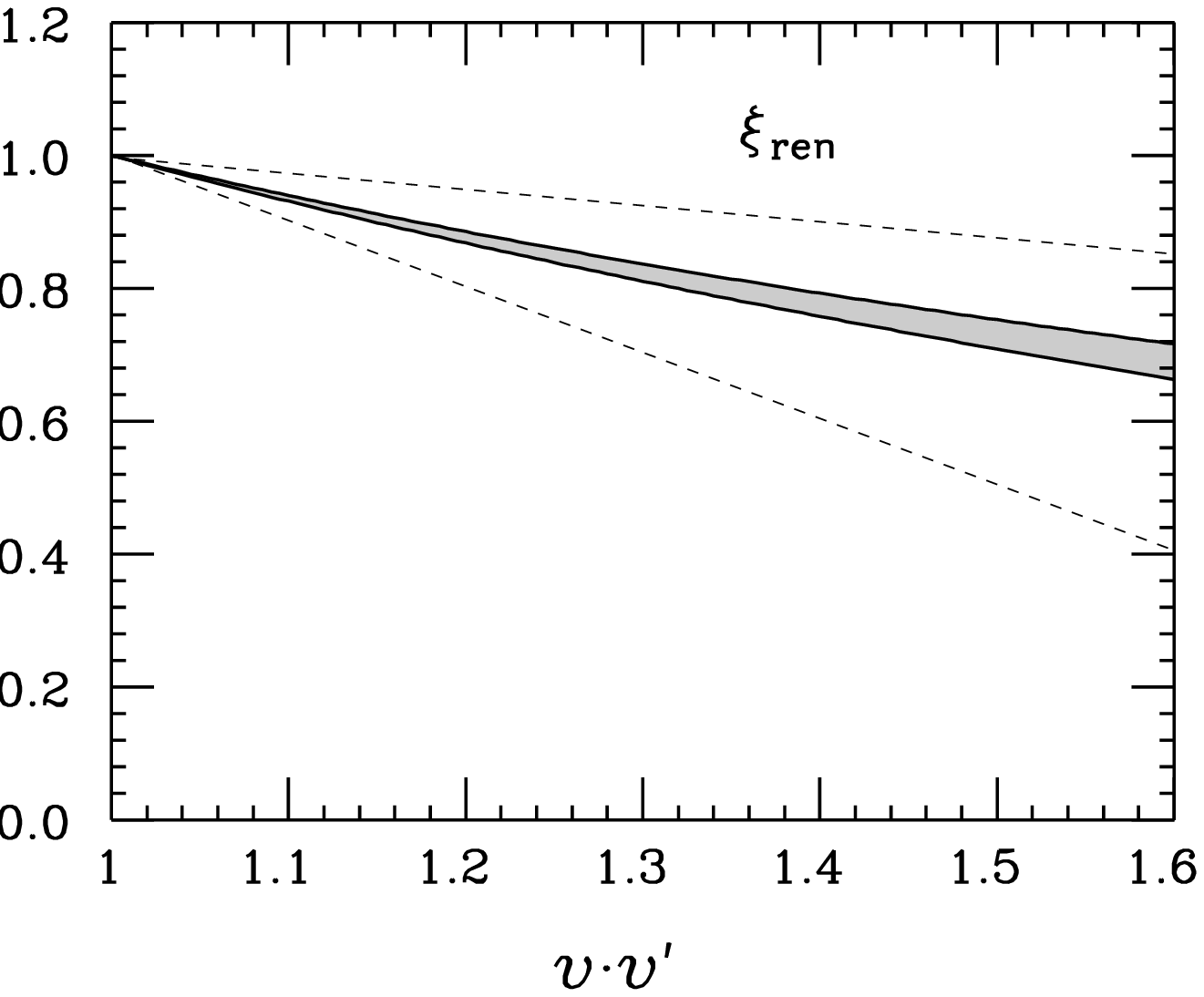}}
\smallskip
\cap{\tenpoint {\bf Figure 4:} Prediction for the Isgur--Wise function.
The dashed lines indicate the Bjorken and Voloshin bounds on the slope
at $y=1$. The shaded region shows only the uncertainty related to
variations of the sum rule parameters (from \mark{\review}).}
\smallskip
\endinsert
\subsection{$1/m_Q$ corrections:}~%
For the four universal functions that appear at order $1/m_Q$ of the
heavy quark expansion, the perturbative order $\alpha_s$ corrections
are even more significant compared to the leading order calculation
\mark{\SRsubl}, than in the case of the two--point function sum rule
\mark{\SRchi,\SRxi}. These corrections can change certain results by
as much as a factor of two.
The form factors $\chi_2$ and $\chi_3$, which parameterize the matrix
elements of the $O_{\rm mag}$ chromomagnetic operator in the effective
Lagrangian, come out to be numerically negligible \mark{\SRchi}.
However, $\xi_3$ is of order unity; it is proportional to the
Isgur--Wise function $\xi(y)$ to a very good approximation \mark{\SRxi}.
It is convenient to present the result in terms of the renormalization
group invariant ratio
$$
\eta(y) \equiv {\xi_3(y,\mu)\over\xi(y,\mu)} = 0.6\pm0.2 \,,\eqno\eq$$
which already includes a conservative estimate of the systematic
uncertainties as well.
This is the result that we have already discussed in Section 2,
which makes the extraction of $|\Vcb|$ surprisingly accurate from
the kinematically suppressed $B\ra D\,\ell\,\bar\nu$ decay mode.

\subsection{In summary:} QCD sum rules provide at present the
theoretically most consistent framework to calculate the universal
functions appearing in HQET, because:
\parskip=0pt
\item\bullet
The Ward identities of the effective theory, \ie\ the zero recoil
conditions are {\it exactly} fulfilled, independent of the sum
rule parameters;
\item\bullet
Due to the simple Feynman rules of HQET, two loop corrections
can be included;
\item\bullet
It allows for systematic renormalization group improvement;
\item\bullet
There is a self--consistency check within the model on the
assumption made.
\par
\noindent \parskip = 4pt plus 1pt minus 1pt
Some of the uncertainties can be estimated in a reliable manner.
These are the ones related to
(a) the truncation of the power series in the OPE (condensates);
(b) higher orders in the perturbative expansion;
(c) numerical values of the condensates.
It is much harder to estimate the systematic uncertainties
inherent in the method, related to the assumption of local duality.
Therefore, it is crucial to determine carefully the sum rule window
and check the stability, which (if found) makes us believe that
duality indeed holds. Due to these systematic uncertainties, the
accuracy of the QCD sum rule method cannot be arbitrarily improved
(unlike the case of lattice calculations).
Experience tells us that if all the above requirements are met,
then the physical predictions of QCD sum rules are accurate to
within~20--30\%.

\section{Form factors}

The predictive power of the above results become more transparent if
we translate them into predictions for the form factors measurable
in $B\ra D^{(*)}\ell\,\bar\nu$ decays. Throughout the following
discussion we use the form factor basis defined in Ref.~\mark{\BSW}.
In this notation, in the vanishing lepton mass limit, $F_1(q^2)$ is
the single measurable form factor in $B\ra D\,\ell\,\bar\nu$ decays,
while $A_1(q^2)$, $A_2(q^2)$, and $V(q^2)$ can be measured in
$B\ra D^{(*)}\ell\,\bar\nu$. In the infinite heavy quark mass limit
\mark{\NR}
$$
\left\(1-{q^2\over(m_B+m_{D^*})^2}\right\)^{-1} A_1(q^2)
= A_2(q^2) = V(q^2) = F_1(q^2) \,.\eqno\eq$$
The finite masses of the $b$ and $c$ quarks introduce symmetry breaking
effects. The predictions for these form factors are plotted in
Ref.~\mark{\review}. It is useful to define ratios of the three form
factors measurable in $B\ra D^*\ell\,\bar\nu$ decays:
$$
R_1(q^2) = \left\(1-{q^2\over(m_B+m_{D^*})^2}\right\)
{V(q^2)\over A_1(q^2)}\,, \qquad
R_2(q^2) = \left\(1-{q^2\over(m_B+m_{D^*})^2}\right\)
{A_2(q^2)\over A_1(q^2)} \,.\eqno\eq$$
Clearly, in the heavy quark symmetry limit $R_1\equiv R_2\equiv1$.
In the following table we present the predictions of our QCD sum rule
analysis in HQET \mark{\SRchi,\SRxi,\review}, together with the quark
models of ISGW \mark{\ISGW}, BSW \mark{\BSW}, KS \mark{\KS}, and QCD
sum rules in ``full" QCD \mark{\PBall} (not including radiative
corrections):
\bigskip
\def\haha{
\rlap{\raise 12pt\hbox{\vbox{\vskip 1pt\noindent HQET +}}}{\rm QCD SR}~~}
\begintable
$B\ra D^*\ell\,\bar\nu$ | \haha | ISGW & BSW  & KS | QCD SR \crthick
$R_1(q^2_{\rm max})$| 1.35 | ~~~1.01~~ & ~~~0.91~~~ & ~~1.09~~~ | 1.31\cr
$R_1(0)$            | 1.27 | 1.27 & 1.09 & 1.00 | 1.23  \crthick
{}~$R_2(q^2_{\rm max})$~| 0.79 | 0.91 & 0.85 & 1.09 | 0.95  \cr
$R_2(0)$            | 0.85 | 1.14 & 1.06 & 1.00 | 1.05
\endtable
\smallskip
\centerline{\tenpoint {\bf Table 2:}
Predictions for the form factor ratios at the endpoins of the spectrum.}
\smallskip
\noindent
The large values of $R_1$ is an almost model--independent prediction of
HQET, in the sense that both the QCD and the $1/m_c$ corrections are
positive, and the $1/m_b$ terms could only cancel the previous two, if
the QCD sum rule prediction for $\xi_3$ failed with an order of
magnitude. On the other hand, $R_2$ depends more strongly on $\xi_3$,
and other models tend to give significantly higher values than ours.
An experimental measurement of this quantity could distinguish
between the above predictions.

\chapter{Summary and Open Problems}

In conclusion, we have seen how heavy quark symmetry reduces hadronic
uncertainties by providing relations among heavy hadron form factors,
and determining the absolute normalization of some of them at the
kinematic limit point. Deviations from the predictions in this symmetry
limit can be taken into account in the framework of a low energy
effective theory. To calculate the universal functions appearing in
HQET, QCD sum rules provide the theoretically most consistent framework.
Of particular interest is the extraction of $|\Vcb|$: the theoretical
uncertainty has been reduced to less than 5\% in the decay
$B\ra D^*\ell\,\bar\nu$, and it should be hardly larger in $B\ra D$
decays. Of course, the experimental measurement of
$B\ra D\,\ell\,\bar\nu$ near zero recoil is more difficult.
However, it would be an independent measurement providing checks on
both the theoretical and the experimental analysis.
The QCD sum rule determination of the universal functions that appear
at order $1/m_Q$ in the heavy quark expansion gives predictions for the
symmetry breaking effects. A measurement of the $R_{1,2}$ form factor
ratios in $B\ra D^*$ decays would distinguish between these sum rule
predictions and other models.

In those cases when the final state does not contain a single
heavy quark, HQS does not yield as restrictive relations as for
heavy to heavy transitions.
In particular, there is no absolute normalization of form factors.
Still, different processes can be related to each other.
This is the idea behind extracting $|\Vub|$ from a comparison of
$B\ra X\,\ell\,\bar\nu$ and $D\ra X\,\ell\,\bar\nu$, where $X$
is either $\pi$ or $\rho$ \mark{\vub}. When $X=\pi$, additional
constraints are provided by chiral symmetry and soft pion theorems.
The main problems are (a) that the contribution of the $B^*$
pole is large, and even dominant in the chiral limit; (b) the $1/m_Q$
terms are expected to be significant. If these difficulties are
overcome then the determination of $|\Vub|$ from the experimental
data may become much more reliable than that at present \mark{\GZMY}.

Here we only discussed exclusive semileptonic decays of heavy mesons.
However, heavy quark symmetry also provides clues for accurate
calculations of inclusive decays \mark{\Adam}. There are also exciting
developments in progress to achieve a reliable, model--independent
understanding of hadronic decays \mark{\Bigi}.
These theoretical developments will hopefully lead to a precise
understanding of heavy hadron physics, ultimately allowing to
distinguish the signatures of new physics in heavy quark systems
(if such signatures exist) from the ever decreasing uncertainties
in the predictions of the Standard Model.

\bigskip
\ack\par
I am grateful to Yossi Nir and Matthias Neubert for a most enjoyable
collaboration on some of the subjects discussed here, and for help
and suggestions in planning this talk.
I thank Arne Freyberger for Fig.~2.
I would like to thank Jonathan Rosner for the invitation;
Gianpaolo Bellini and the organizers for support during my stay at Pavia.

\endpage
\baselineskip=14pt
\refout
\end